\documentclass[twoside]{article} 
\usepackage{qic,epsfig} 
\usepackage{amsmath}
\usepackage{cite}

\newcommand{\calG}{\mathcal{G}}
\newcommand{\calQ}{\mathcal{Q}}
\newcommand{\calH}{\mathcal{H}}
\newcommand{\calF}{\mathcal{F}}
\newcommand{\calN}{\mathcal{N}}

\newcommand{\calGU}{\mathcal{G}_{U}} 
\newcommand{\bfF}{\mathbf{F}}
\newcommand{\bfsig}{\mbox{\boldmath{$\sigma$}}}
\newcommand{\bfhatz}{\hat{\mathbf{z}}} 
\newcommand{\bfhatx}{\hat{\mathbf{x}}}
\newcommand{\bfhaty}{\hat{\mathbf{y}}}

\renewcommand{\thefootnote}{\fnsymbol{footnote}} 

\begin{document} 
\setlength{\textheight}{8.0truein}    

\runninghead{High-fidelity universal quantum gates}
           {Li and Gaitan}

\normalsize\textlineskip 
\thispagestyle{empty} 
\setcounter{page}{1}


\vspace*{0.88truein}

\alphfootnote

\fpage{1}

\centerline{\bf 
HIGH-FIDELITY UNIVERSAL QUANTUM GATES}
\vspace*{0.035truein} 
\centerline{\bf THROUGH GROUP-SYMMETRIZED RAPID PASSAGE}
\vspace*{0.37truein} 
\centerline{\footnotesize RAN LI} 
\vspace*{0.015truein}
\centerline{\footnotesize\it Department of Physics, Kent State University,
Stark Campus} 
\baselineskip=10pt 
\centerline{\footnotesize\it North Canton, OH 44720;} 
\vspace*{0.015truein}
\centerline{\footnotesize\it Advanced Sciences Institute, The Institute of
Physical and Chemical Research (RIKEN)} 
\baselineskip=10pt
\centerline{\footnotesize\it Wako-shi, Saitama 351-0198, Japan;}
\vspace*{0.015truein} 
\centerline{\footnotesize\it CREST, Japan Science and Technology Agency 
(JST)} 
\baselineskip=10pt 
\centerline{\footnotesize\it Kawaguchi, Saitama 332-0012, Japan} 
\vspace*{10pt} 
\centerline{\footnotesize FRANK GAITAN\footnote{Corresponding author: 
fgaitan@lps.umd.edu}  \footnote{Permanent Address: Laboratory for Physical 
Sciences, 8050 Greenmead Drive, College Park, MD 20740.}}
\vspace*{0.015truein} 
\centerline{\footnotesize\it Advanced Sciences Institute, The Institute of 
Physical and Chemical Research (RIKEN)} 
\baselineskip=10pt 
\centerline{\footnotesize\it Wako-shi, Saitama 351-0198, Japan;} 
\vspace*{0.015truein} 
\centerline{\footnotesize\it CREST, Japan Science and Technology Agency 
(JST)} 
\baselineskip=10pt
\centerline{\footnotesize\it Kawaguchi, Saitama 332-0012, Japan}
\vspace*{0.225truein} 

\vspace*{0.21truein}

\abstracts{ Twisted rapid passage is a type of non-adiabatic rapid passage 
that generates controllable quantum interference effects that were first 
observed experimentally in $2003$. It is shown that twisted rapid passage 
sweeps can be used to implement a universal set of quantum gates $\calGU$ 
that operate with high-fidelity. The gate set $\calGU$ consists of the 
Hadamard and NOT gates, together with variants of the phase, $\pi /8$, and 
controlled-phase gates. For each gate $g$ in $\calGU$, sweep parameter 
values are provided which simulations indicate will produce a unitary operation
that approximates $g$ with error probability$P_{e} < 10^{-4}$. Note that 
\textit{all\/} gates in $\calGU$ are implemented using a \textit{single
family\/} of control-field, and the error probability for each gate falls below 
the rough-and-ready estimate for the accuracy threshold $P_{a}\sim 
10^{-4}$.}{}{}

\vspace*{10pt}

\vspace*{3pt} 

\vspace*{1pt}\textlineskip 
\vspace*{-0.5pt} 
\noindent 

\setcounter{footnote}{0}
\renewcommand{\thefootnote}{\alph{footnote}}

\section{Introduction} 
\label{sec1}

The accuracy threshold theorem \cite{ft1,ft2,ft3,ft4,ft5,ft6,ft7,ft8} provides
the impetus for the work presented in this paper. This remarkable theorem
established that an arbitrary quantum computation could be done with an
arbitrarily small error probability, in the presence of noise, and using
imperfect quantum gates, if the following conditions are satisfied. (1)~The
computational data is protected by a sufficiently layered concatenated 
quantum error correcting code. (2)~Fault-tolerant protocols for quantum 
computation, error correction, and measurement are used. (3)~A universal 
set of \textit{unencoded\/} quantum gates is available with the property that 
each gate in the set has an error probability $P_{e}$ that falls below a value
$P_{a}$ known as the accuracy threshold. The value of the threshold is
model-dependent, though for many, $P_{a}\sim 10^{-4}$ has become a
rough-and-ready estimate. Thus gates are anticipated to be approaching the
accuracies needed for fault-tolerant quantum computing when $P_{e}< 
10^{-4}$. One of the principal challenges facing the field of quantum 
computing is determining how to implement a universal set of unencoded 
quantum gates for which all gate error probabilities satisfy $P_{e}<10^{-4}$.

In this paper numerical simulation results are presented which indicate that a
class of non-adiabatic rapid passage sweeps, first realized experimentally in
1991 \cite{zw1}, and known as twisted rapid passage (TRP), should be 
capable of implementing a universal set of unencoded quantum gates $\calGU$ 
that operate non-adiabatically and with gate error probabilities satisfying 
$P_{e}< 10^{-4}$. The gate set $\calGU$ consists of the one-qubit Hadamard 
and NOT gates, together with variants of the one-qubit $\pi /8$ and phase 
gates, and the two-qubit controlled-phase gate. The universality of $\calGU$ 
was demonstrated in Ref.~\cite{lhg2}. This level of gate accuracy is largely 
due to controllable quantum interference effects that arise during a TRP 
sweep \cite{fg1,lhg1}, and which were observed in 2003 using NMR 
\cite{zwan2}. To find TRP sweep parameter values that yield such 
high-performance quantum gates, it proved necessary to combine numerical 
simulation of the Schrodinger dynamics with optimization algorithms that search
for minima of $P_{e}$. In the case of the two-qubit modified controlled-phase 
gate, to achieve $P_{e}< 10^{-4}$, it was also necessary to interleave the 
TRP sweep with the group-symmetrized evolution of Ref.~\cite{zan}.

The outline of this paper is as follows. Following this Introduction,
Section~\ref{sec2} summarizes the essential properties of TRP, and
how the numerical simulation and optimization are done. In
Section~\ref{sec3} we explain how group-symmetrized evolution is 
incorporated into a TRP sweep, and then present our simulation results for 
each of the gates in $\calGU$ in Section~\ref{sec4}. We close in 
Section~\ref{sec5} with a summary of our results, and a discussion of 
possible directions for future work.

\section{Twisted Rapid Passage} 
\label{sec2} 

\noindent This Section presents the essential properties of twisted rapid 
passage (TRP). A more detailed presentation of the discussion in
Sections~\ref{sec2.1} and \ref{sec2.2} appears in Refs.~\cite{lhg2}
and \cite{lhg1}. In an effort to make the present paper more self-contained, 
we summarize that discussion here. Section~\ref{sec2.1}  introduces TRP 
as a generalization of adiabatic rapid passage in which the control-field twists 
in the $x$-$y$ plane at the same time that its $z$-component undergoes a 
non-adiabatic inversion.  We describe how controllable quantum interference 
effects arise as a consequence of the twisting. Section~\ref{sec2.2} discusses 
the details of the numerical simulations, and describes the optimization 
procedures used to find minima of the gate error probability\footnote{As will 
be seen shortly, we actually search for minima of an upper bound of the gate 
error probability.}.

\subsection{TRP Essentials} 
\label{sec2.1} 

\noindent To introduce TRP \cite{fg1,lhg1}, we consider a single-qubit 
interacting with an external control-field $\bfF (t)$ via the Zeeman
interaction $H_{z}(t) = -\bfsig\cdot\bfF (t)$, where $\sigma_{i}$ are 
the Pauli matrices ($i=x,y,z$). TRP is a generalization of adiabatic rapid 
passage (ARP) \cite{abra}. In ARP, the control-field $\bfF (t)$ is slowly 
inverted over a time $T_{0}$ such that $\bfF (t) =at\,\bfhatz + b\,\bfhatx$. 
In TRP, however, the control-field is allowed to twist in the $x$-$y$ plane 
with time-varying azimuthal angle $\phi (t)$, while simultaneously 
undergoing inversion along the $z$-axis:
\begin{equation}
\bfF (t) = at\,\bfhatz + b\,\cos\phi (t)\,\bfhatx +b\,\sin\phi (t)\,\bfhaty .
\label{TRPsweep} 
\end{equation} 
Here $-T_{0}/2\leq t\leq T_{0}/2$, and throughout this paper, we consider 
TRP with \textit{non-adiabatic\/} inversion. As shown in Ref.~\cite{lhg1}, 
the qubit undergoes resonance when 
\begin{equation} 
at -\frac{\hbar}{2}\frac{d\phi}{dt} = 0 . 
\label{rescon}
\end{equation} 
For polynomial twist, the phase profile $\phi (t)$ takes the form 
\begin{equation} 
\phi_{n}(t) = \frac{2}{n}Bt^{n} . 
\label{polytwist} 
\end{equation} 
In this case, Eq.~(\ref{rescon}) has $n-1$ roots, though only real-valued roots 
correspond to resonance. Ref.~\cite{fg1} showed that for $n\geq 3$, the 
qubit undergoes resonance multiple times during a \textit{single\/} TRP sweep: 
(i)~for all $n\geq 3$, when $B>0$; and (ii)~for odd $n\geq 3$, when $B<0$. 
For the remainder of this paper we restrict ourselves to $B>0$, and to 
\textit{quartic\/} twist for which $n=4$ in Eq.~(\ref{polytwist}). During 
quartic twist, the qubit passes through resonance at times $t=0,\pm
\sqrt{a/\hbar B}$ \cite{fg1}. It is thus possible to vary the time separating 
the resonances by varying the TRP sweep parameters $B$ and $a$.

Ref.~\cite{fg1} showed that these multiple resonances have a strong 
influence on the qubit transition probability, allowing transitions to be
strongly enhanced or suppressed through a small variation of the sweep 
parameters. Ref.~\cite{fg2} calculated the qubit transition amplitude to all
orders in the non-adiabatic coupling. The result found there can be 
re-expressed as the following diagrammatic series: 
\begin{equation} 
\setlength{\unitlength}{0.05in}
T_{-}(t) = \begin{picture}(10,4) 
                       \put(10,-1.5){\vector(-1,0){3.25}}
                       \put(5,-1.5){\line(1,0){1.75}} 
                       \put(5,-1.5){\vector(0,1){3.25}}
                       \put(5,1.75){\line(0,1){1.75}} 
                       \put(5,3.5){\vector(-1,0){3.25}}
                       \put(0,3.5){\line(1,0){1.75}} 
                 \end{picture} 
\hspace{0.05in} +
      \begin{picture}(20,4) 
           \put(20,-1.5){\vector(-1,0){3.25}}
           \put(15,-1.5){\line(1,0){1.75}} 
           \put(15,-1.5){\vector(0,1){3.25}}
           \put(15,1.75){\line(0,1){1.75}} 
           \put(15,3.5){\vector(-1,0){3.25}}
           \put(10,3.5){\line(1,0){1.75}} 
           \put(10,3.5){\vector(0,-1){3.25}}
           \put(10,-1.5){\line(0,1){1.75}} 
           \put(10,-1.5){\vector(-1,0){3.25}}
           \put(5,-1.5){\line(1,0){1.75}} 
           \put(5,-1.5){\vector(0,1){3.25}}
           \put(5,1.75){\line(0,1){1.75}} 
           \put(5,3.5){\vector(-1,0){3.25}}
           \put(0,3.5){\line(1,0){1.75}} 
     \end{picture} 
\hspace{0.05in} +
           \begin{picture}(30,5)
              \put(30,-1.5){\vector(-1,0){3.25}}
              \put(25,-1.5){\line(1,0){1.75}}
              \put(25,-1.5){\vector(0,1){3.25}}
              \put(25,1.75){\line(0,1){1.75}}
              \put(25,3.5){\vector(-1,0){3.25}}
              \put(20,3.5){\line(1,0){1.75}}
              \put(20,-1.5){\vector(-1,0){3.25}}
              \put(20,3.5){\vector(0,-1){3.25}}
              \put(20,-1.5){\line(0,1){1.75}}
              \put(15,-1.5){\line(1,0){1.75}}
              \put(15,-1.5){\vector(0,1){3.25}}
              \put(15,1.75){\line(0,1){1.75}}
              \put(15,3.5){\vector(-1,0){3.25}}
              \put(10,3.5){\line(1,0){1.75}}
              \put(10,3.5){\vector(0,-1){3.25}}
              \put(10,-1.5){\line(0,1){1.75}}
              \put(10,-1.5){\vector(-1,0){3.25}}
              \put(5,-1.5){\line(1,0){1.75}}
              \put(5,-1.5){\vector(0,1){3.25}}
              \put(5,1.75){\line(0,1){1.75}}
              \put(5,3.5){\vector(-1,0){3.25}}
              \put(0,3.5){\line(1,0){1.75}}
           \end{picture}
\hspace{0.1in} + \hspace{0.1in} \cdots \hspace{0.1in} . 
\label{diagser} 
\end{equation} 
Lower (upper) lines correspond to propagation in the negative (positive) 
energy-level, and the vertical lines correspond to transitions between the 
two energy-levels. The calculation sums the probability amplitudes for all
interfering alternatives \cite{f&h} that allow the qubit to end up in the 
positive energy-level given that it was initially in the negative energy-level.
As we have seen, varying the TRP sweep parameters varies the time 
separating the resonances. This in turn changes the value of each diagram 
in Eq.~(\ref{diagser}), and thus alters the interference between the 
alternative transition pathways. It is the sensitivity of the individual
alternatives/diagrams to the time separation of the resonances that 
allows TRP to manipulate this quantum interference. Zwanziger et al.\ 
\cite{zwan2} observed these interference effects in the transition 
probability using NMR and found quantitative agreement between theory 
and experiment. It is this link between interfering quantum alternatives 
and the TRP sweep parameters that we believe underlies the ability of 
TRP to drive high-fidelity (non-adiabatic) one- and two-qubit gates.

\subsection{Simulation and Optimization Procedures} 
\label{sec2.2} 
\noindent As is well-known, the Schrodinger dynamics is driven by a
Hamiltonian $H(t)$ which causes a unitary transformation $U$ to be 
applied to an initial quantum state $|\psi\rangle$. In this paper, it is 
assumed that the Hamiltonian $H(t)$ contains terms that Zeeman-couple 
each qubit to the TRP control-field $\bfF (t)$. Assigning values to the 
TRP sweep parameters $(a,b,B,T_{0})$ fixes the control-field $\bfF (t)$, 
and in turn, the actual unitary transformation $U_{a}$ applied to 
$|\psi\rangle$. The task  is to find TRP sweep parameter values that
produce an applied gate $U_{a}$ that approximates a desired target 
gate $U_{t}$ sufficiently closely that its error probability (defined below) 
satisfies $P_{e}<10^{-4}$. In the following, the target gate $U_{t}$ 
will be one of the gates in the universal set $\calGU$. Since $\calGU$ 
contains only one- and two-qubit gates, our simulations will only involve 
one- and two-qubit systems.

For the \textit{one-qubit simulations}, the Hamiltonian $H_{1}(t)$ is the 
Zeeman Hamiltonian $H_{z}(t)$ introduced in Section~\ref{sec2.1}. 
Ref.~\cite{lhg1} showed that it can be written in the following 
dimensionless form:
\begin{equation}
\calH_{1} (\tau ) = (1/\lambda)\,\left\{ -\tau\sigma_{z} 
                                 -\cos\phi_{4}(\tau )\sigma_{x} 
                                   -\sin\phi_{4}(\tau )\sigma_{y}\right\} .
\label{oneqbtHam}
\end{equation}
Here: $\tau = (a/b)t$; $\lambda = \hbar a/b^{2}$; and for quartic twist,
$\phi_{4}(\tau ) = (\eta_{4}/2\lambda )\tau^{4}$, with $\eta_{4}=\hbar 
Bb^{2}/a^{3}$.

For the \textit{two-qubit simulations}, the Hamiltonian $H_{2}(t)$ 
contains terms that Zeeman-couple each qubit to the TRP control-field, 
and an Ising interaction term that couples the two qubits. Alternative 
two-qubit interactions can easily be considered, though all simulation 
results presented below assume an Ising interaction between the qubits. 
To break a resonance-frequency degeneracy $\omega_{12}=\omega_{34}$
for transitions between, respectively, the ground and first-excited states 
($E_{1}\leftrightarrow E_{2}$) and the second- and third excited states 
($E_{3}\leftrightarrow E_{4}$), the term $c_{4}|E_{4}(t )\rangle\langle 
E_{4}(t )|$ was added to $H_{2}(t)$. Combining all of these remarks, we 
arrive at the following (dimensionless) two-qubit Hamiltonian \cite{lhg2}:
\begin{eqnarray}
\calH_{2}(\tau ) & = & 
               \left[ -(d_{1}+d_{2})/2+\tau/\lambda\right]\sigma_{z}^{1} 
                  -(d_{3}/\lambda )\left[\cos\phi_{4}\sigma_{x}^{1} +
                      \sin\phi_{4}\sigma_{y}^{1}\right]\nonumber\\
   & &     \hspace{0.275in}       +\left[ -d_{2}/2+\tau/\lambda\right]
                   \sigma_{z}^{2} 
             \hspace{0.075in}  -  (1/\lambda )\left[\cos\phi_{4}\sigma_{x}^{2} 
                   +       \sin\phi_{4}\sigma_{y}^{2}\right]\nonumber\\
   & &     \hspace{0.55in}        -(\pi d_{4}/2)\sigma_{z}^{1}\sigma_{z}^{2} 
                  \hspace{0.1in} +c_{4}|E_{4}(\tau )\rangle\langle 
                        E_{4}(\tau )| .
\label{twoqbtHam}
\end{eqnarray}
Here: (i)~$b_{i} = \hbar\gamma_{i}B_{rf}/2$, $\omega_{i}=\gamma_{i}
B_{0}$, $\gamma_{i}$ is the gyromagnetic ratio for qubit $i$, and $i=1,2$; 
(ii)~$\tau = (a/b_{2})t$, $\lambda = \hbar a/b_{2}^{2}$, and $\eta_{4}=
\hbar Bb_{2}^{2}/a^{3}$; and (iii)~$d_{1}=(\omega_{1}-\omega_{2})
b_{2}/a$, $d_{2}=(\Delta /a)b_{2}$, $d_{3} = b_{1}/b_{2}$, and 
$d_{4}=(J/a)b_{2}$, where $\Delta$ is a detuning parameter \cite{lhg2}.

The numerical simulations assign values to the TRP sweep parameters and 
then integrate the Schrodinger equation to obtain the unitary transformation
$U_{a}$ produced by the resulting TRP sweep. Given $U_{a}$, $U_{t}$, and the 
initial state $|\psi\rangle$, it is possible to work out \cite{lhg1} the error 
probability $P_{e}(\psi )$ for the TRP final state $|\psi_{a}\rangle =
U_{a}|\psi\rangle$, relative to the target final state $|\psi_{t}\rangle 
= U_{t}|\psi\rangle$. The gate error probability $P_{e}$ is defined to be 
the worst-case value of $P_{e}(\psi )$:
\begin{equation}
P_{e}\equiv \max_{|\psi\rangle}P_{e}(\psi ) .
\label{errprobdef}
\end{equation}
It proves useful at this point to introduce the positive operator
\begin{equation}
P = \left( U_{a}^{\dagger}-U_{t}^{\dagger}\right)\left( U_{a}-U_{t}
        \right) .
\label{Pdef}
\end{equation}
Ref.~\cite{lhg1} showed that the error probability $P_{e}$ satisfies the 
upper bound 
\begin{equation}
P_{e}\leq Tr\, P,
\label{upprbnd}
\end{equation} 
where the RHS is the trace of the positive operator $P$. Once $U_{a}$ 
is known, $Tr\, P$ is easily evaluated, and so it is a convenient proxy 
for $P_{e}$ which is harder to calculate. $Tr\, P$ also has the virtue of 
being directly related to the gate fidelity
\begin{equation}
\calF_{n} = \left(\frac{1}{2^{n}}\right)\, Re\left[\, Tr\left( 
                     U_{a}^{\dagger}U_{t}\right)\,\right] ,
\label{gfiddef}
\end{equation}
where $n$ is the number of qubits acted on by the gate. It is 
straightforward to show \cite{lhg2} that 
\begin{equation}
\calF_{n} = 1 -\left(\frac{1}{2^{n+1}}\right)\, Tr\, P .
\label{fidtrP}
\end{equation}
The simulations calculate $Tr\, P$, which is then used to upper bound 
the gate error probability $P_{e}$ via Eq.~(\ref{upprbnd}).

To find TRP sweep parameter values that yield highly accurate non-adiabatic
quantum gates, it proved necessary to combine the numerical simulations with
function minimization algorithms \cite{numrec} that search for sweep parameter 
values that minimize the $Tr\, P$ upper bound. The multi-dimensional
downhill simplex method was used for the one-qubit gates, while simulated
annealing was used for the two-qubit modified controlled-phase gate. This
produced the one-qubit gate results that will be presented in 
Section~\ref{sec4.1}. However, for the modified controlled-phase gate,
simulated annealing was only able to find sweep parameter values that
gave $P_{e}\leq 1.27\times 10^{-3}$ \cite{lhg2}. To further improve
the performance of this two-qubit gate, it proved necessary to incorporate
the group-symmetrized evolution of Ref.~\cite{zan} to obtain a modified
controlled-phase gate with $P_{e}< 10^{-4}$. In the following Section we 
describe how group-symmetrized evolution is incorporated into a TRP
sweep. 

\section{Group-symmetrized Evolution and TRP} 
\label{sec3} 
\noindent Ref.~\cite{zan} introduced a unitary group-symmetrization
procedure that yields an effective dynamics that is invariant under the
action of a finite group $\mathcal{G}$. To incorporate this  
group-symmetrization into a TRP sweep, the first step is to identify the 
group $\mathcal{G}$ with a finite symmetry group of the target gate 
$U_{t}$, and then apply the procedure of Ref.~\cite{zan} to filter out 
the $\mathcal{G}$-noninvariant part of the TRP dynamics. As the
$\mathcal{G}$-noninvariant dynamics is manifestly bad dynamics
relative to $U_{t}$, group-symmetrized TRP yields a better approximation
to $U_{t}$. In this section we describe how group-symmetrization works, 
and then show how it can be incorporated into a TRP sweep. The simulation 
results presented in Section~\ref{sec4.2} for the two-qubit modified 
controlled-phase gate are for a group-symmetrized TRP sweep.

\subsection{Static Hamiltonian}
\label{sec3.1}
\noindent Consider a quantum system $\calQ$ with time-independent 
Hamiltonian $H$ and Hilbert space $\calH$. The problem is to provide 
$\calQ$ with an effective dynamics that is invariant under a finite group 
$\calG$, even when $H$ itself is \textit{not\/} $\calG$-invariant. This 
symmetrized dynamics manifests as a $\calG$-invariant effective propagator 
$\tilde{U}$ that evolves the system state over a time $t$. Let $\{\rho_{i} = 
\rho (g_{i})\}$ be a unitary representation of $\calG$ on $\calH$, and let 
$|\calG |$ denote the order of $\calG$. The procedure begins by 
partitioning the time-interval $(0,t)$ into $N$ subintervals of duration 
$\Delta t_{N} = t/N$, and then further partioning each subinterval into 
$|\calG |$ smaller intervals of duration $\delta t_{N} = \Delta t_{N}/
|\calG |$. Let $\delta U_{N} = \exp\left[ -(i/\hbar )\delta t_{N} H\right]$ 
denote the $H$-generated propagator for a time-interval $\delta t_{N}$, 
and assume that the time to apply each $\rho_{i}\in\calG$ is negligible 
compared to $\delta t_{N}$ (bang-bang limit \cite{vio&llyd}). In each 
subinterval, the following sequence of transformations is applied:
\begin{equation}
U(\Delta t_{N}) = \prod_{i=1}^{|\calG |} \rho_{i}^{\dagger}\delta U_{N}
                               \rho_{i} .
\label{smallprop}
\end{equation}
The propagator $\tilde{U}$ over the full time-interval $(0,t)$ is then
\begin{equation}
\tilde{U} = \lim_{N\rightarrow\infty}\left[ U(\Delta t_{N}\right]^{N}.
\end{equation}
Ref.~\cite{zan} showed that:
\begin{enumerate}
\item for $N\gg 1$, $U(\Delta t_{N})\rightarrow \exp\left[ -(i/\hbar )
\Delta t_{N}\tilde{H}\right]$, where $\tilde{H} = (1/|\calG |)
\sum_{i=1}^{|\calG |} \rho_{i}^{\dagger} H\rho_{i}$;
\item $\tilde{H}$ is $\calG$-invariant (viz.\ $[\tilde{H},\rho_{i}]=0$
for all $\rho_{i}\in\calG$);
\item the propagator $\tilde{U}$ over $(0,t)$ is $\tilde{U}=\exp\left[
-(i/\hbar )t\tilde{H}\right]$, which is $\calG$-invariant due to the 
$\calG$-invariance of $\tilde{H}$.
\end{enumerate}
The outcome of this procedure is an effective propagator $\tilde{U}$ 
that is $\calG$-invariant as desired.

\subsection{Time-varying Hamiltonian}
\label{sec3.2}
\noindent In this subsection we show how the above procedure can be 
generalized to allow for a time-varying Hamiltonian $H(t)$ which is the 
appropriate context for a TRP sweep. Although the generalization is 
straight-forward, we are not aware of any prior treatment of 
group-symmetrized evolution in the presence of a time-dependent Hamiltonian. 
The essential idea is to divide the time interval $(0,t)$ into sufficiently 
small subintervals that $H(t)$ is effectively constant in each. The above 
time-independent argument can be then be applied to each subinterval, 
yielding a $\calG$-symmetrized propagator for that subinterval. Time-ordering 
the subinterval effective propagators then gives the effective group-symmetrized 
propagator for the entire interval $(0,t)$. Having sketched out the basic 
idea, we now present the details. 

Consider a quantum system $\calQ$ evolving under the action of a 
time-varying Hamiltonian $H(t)$. We begin again by partitioning the
time-interval $(0,t)$ into $\calN$ subintervals $(t_{i-1},t_{i})$ of 
duration $\Delta t_{\calN}=t/\calN$, where $i=1,\ldots ,\calN$. The number
of subintervals is chosen sufficiently large that $H(t)$ is effectively 
constant over each subinterval: $H(t)\approx H(t_{i})$ for all $t\in
(t_{i-1},t_{i})$. We estimate the value of $\calN$ for a TRP sweep in
Section~\ref{sec3.3}. Note that for a static Hamiltonian, this requirement
is true for all values of $\calN$. Since $H(t)\approx H(t_{i})$ in the $i^{th}$
subinterval $(t_{i-1},t_{i})$, we can apply the symmetrization procedure
of Section~\ref{sec3.1} to this subinterval. The result is the 
$\calG$-invariant effective propagator
\begin{equation}
\tilde{U}(t_{i},t_{i-1}) = \exp\left[ -(i/\hbar )\Delta t_{\calN}
                                          \tilde{H}(t_{i})\right],
\end{equation}
where
\begin{equation}
\tilde{H}(t_{i}) = \frac{1}{|\calG |}\sum_{i=1}^{|\calG |}\rho_{i}^{\dagger}
                                 H(t_{i})\rho_{i}
\end{equation}
is the $\calG$-invariant effective Hamiltonian for $(t_{i-1},t_{i})$. Having
the effective propagator for each subinterval, the effective group-symmetrized 
propagator $\tilde{U}$ for the entire time-interval $(0,t)$ is simply the 
time-ordered product of the $\tilde{U}(t_{i},t_{i-1})$:
\begin{equation}
\tilde{U} = T\left[\, \exp\left( -i/\hbar\int_{0}^{t}\, d\tau \tilde{H}(\tau )
                   \right)\right] ,
\label{fullprop}
\end{equation}
where $T$ indicates a time-ordered exponential, and
\begin{equation}
\tilde{H}(t) = \frac{1}{|\calG |}\sum_{i=1}^{|\calG |}\rho_{i}^{\dagger}
                         H(t)\rho_{i} .
\label{lclsymHam}
\end{equation}
As in Section~\ref{sec3.1}, we assume the group-symmetrizing pulses 
$\rho_{i}$ can be applied in a time that is much less than $\Delta t_{\calN}/
|\calG |$.

\subsection{TRP}
\label{sec3.3}
\noindent We now describe how group-symmetrized evolution can be 
incorporated into a TRP sweep. For our two-qubit simulations, the target gate 
is the modified controlled-phase gate
\begin{equation}
V_{cp} = \left(\frac{1}{2}\right)\left[\left( I^{1}+\sigma_{z}^{1}\right) I^{2}
                 -\left( I^{1}-\sigma_{z}^{1}\right)\sigma_{z}^{2}\right]
\label{modcfaz}
\end{equation}
which is invariant under the group $\calG = \{ I^{1}I^{2},\sigma_{z}^{1} I^{2},
I^{1}\sigma_{z}^{2},\sigma_{z}^{1}\sigma_{z}^{2}\}$. Thus $|\calG |=4$, and 
we set $\rho_{1}=I^{1}I^{2},\ldots , \rho_{4}=\sigma_{z}^{1}\sigma_{z}^{2}$.
Switching over to dimensionless time, we partition the sweep time-interval
$(-\tau_{0}/2,\tau_{0}/2)$ into $\calN$ subintervals of duration 
$\Delta t_{\calN} = \tau_{0}/\calN$. We want $\calN$ to be sufficiently large
that the two-qubit Hamiltonian $H_{2}(\tau )$ is effectively constant within 
each subinterval. Since the interference effects arise from the twisting of the 
control field, we estimate the size of $\calN$ by requiring that the angle
$\Delta \phi$ swept through by the control field in a time $\Delta t_{\calN}$
is small compared to the final twist angle $\phi_{f}=\phi (\tau_{0}/2)$. 
Specifically, if we require that $\Delta\phi /\phi_{f}<5\times 10^{-3}$, and 
noting that $\Delta\phi (\tau )\approx \dot{\phi}(\tau )\Delta t_{\calN}$ is 
largest at $\tau=\tau_{0}$, it follows that
\begin{eqnarray*}
\Delta t_{\calN} & = & \Delta\phi /\dot{\phi} \\
  & < & 0.005[\phi_{f}/\dot{\phi}(\tau_{0}/2)]\\ 
  & < & 7.5\times 10^{-2}.
\end{eqnarray*}
Recalling that $\Delta t_{\calN} =\tau_{0}/\calN$, and noting that 
$\tau_{0}=120$ for the two-qubit simulations (see Section~\ref{sec4.2}), 
one finds that
\begin{displaymath}
\calN > 1600.
\label{intest}
\end{displaymath}
We shall see in Section~\ref{sec4.2} that $\calN = 2500$ in the two-qubit
simulations which produces an even smaller $\Delta\phi /\phi_{f}$, and so 
enhances the approximation of a constant $H_{2}(\tau )$ in each subinterval. 

The two-qubit numerical simulations for group-symmetrized TRP partition the 
sweep interval $(-\tau_{0}/2,\tau_{0}/2)$ into $\calN$ subintervals as just 
described. Each subinterval $(t_{i-1},t_{i})$ is further partitioned into 
$|\calG |=4$ sub-subintervals by introducing intermediate times $t_{i,j} 
\equiv t_{i-1}+j(\Delta t_{\calN}/4)$, with $j=1,\ldots , 4$. The numerical 
integration over the $i^{th}$ subinterval begins by driving the quantum state 
using a TRP sweep from $t_{i-1}\rightarrow t_{i,1}$, at which time the 
group-symmetrizing pulse $\rho_{2}\rho_{1}^{\dagger}$ is applied. The 
integration then resumes with TRP evolving the state from $t_{i,1}\rightarrow 
t_{i,2}$, followed by application of $\rho_{3}\rho_{2}^{\dagger}$ at time 
$t_{i,2}$. This alternation of TRP-driven evolution and group-symmetrizing 
pulses continues until the final time $t_{i}$ is reached. The numerical 
integration begins at $-\tau_{0}/2$ and continues 
across the subintervals until the final time $\tau_{0}/2$ is reached. Note 
that the time required to apply the group-symmetrizing pulses is assumed to 
be much shorter than $\Delta t_{\calN}/4$. These pulses are applied in the
simulation as matrix operations on the state. This is consistent with our 
bang-bang assumption, although it precludes a study of the effects of
pulse imperfections on gate performance. This limitation will be removed in 
future work. Completion of the numerical integration yields the 
$\calG$-symmetrized TRP propagator $\tilde{U}$ corresponding to a particular 
choice of the sweep parameters $(\lambda , \eta_{4}, \tau_{0})$ and system 
parameters $(c_{4},d_{1},\ldots ,d_{4})$. $\tilde{U}$ then serves as $U_{a}$ 
in the simulated annealing optimization which returns optimized values for the 
sweep and system parameters, as well as the group-symmetrized gate 
$\tilde{U}_{a}$ that best approximates the modified controlled-phase gate 
$V_{cp}$. To reduce the size of the search space, only $(\eta_{4}, c_{4},
d_{1}, d_{4})$ are optimized; $(\tau_{0},\lambda , d_{2},d_{3})$ are 
assigned values. The results of this simulation/symmetrization/optimization 
procedure appear in Section~\ref{sec4.2}. We shall see below that 
$\calG$-symmetrized TRP yields an approximation to $V_{cp}$ with 
$P_{e}< 10^{-4}$.

\section{Simulation Results} 
\label{sec4} 
\noindent Here we present our simulation results for the TRP-driven 
approximations to the gates in the universal set $\calGU$.
The one-qubit gates are discussed in Section~\ref{sec4.1}, while
the two-qubit modified controlled-phase gate appears in
Section~\ref{sec4.2}.

\subsection{One-qubit Gates} 
\label{sec4.1} 
\noindent Operator expressions for the one-qubit target gates in $\calGU$
are:
\begin{displaymath}
\begin{array}{rl}
\mathit{Hadamard:\rule[0ex]{0cm}{2ex}} & U_{h}=(1/\sqrt{2})
     \left(\, \sigma_{z}+     \sigma_{x}\right) ;\\
\mathit{NOT:\rule[0.5ex]{0cm}{2ex}} & U_{not} = \sigma_{x};\\
\mathit{Modified\; \mbox{$\pi /8$}:\rule[0.5ex]{0cm}{2ex}} & 
    V_{\pi /8} = \cos\left( \pi /8\right)\,\sigma_{x} -\sin\left(\pi /8
      \right)\,\sigma_{y};\\
\mathit{Modified\: phase:\rule[0.5ex]{0cm}{2ex}} & V_{p} = 
   (1/\sqrt{2})\left(\, \sigma_{x}      -\sigma_{y}\,\right) .
\end{array}
\end{displaymath}
A study of the TRP-implementation of these gates was first reported in 
Ref.~\cite{lhg1}. The essential results are included here for the reader's
convenience, though space limitations do not allow inclusion of the unitary
operator $U_{a}$ associated with each gate. The interested reader can find
them displayed in Ref.~\cite{lhg1}. The connection between the TRP
experimental and theoretical parameters is given in Refs.~\cite{lhg2} and
\cite{lhg1} for superconducting and NMR qubits, respectively.

Table~\ref{table1}
\begin{table}[hb]
\tcaption{\label{table1}Simulation results for the one-qubit gates in $\calGU$.
The error probability for each gate satisfies $P_{e}\leq Tr\, P$.}
\centerline{\footnotesize\smalllineskip
\begin{tabular}{|c|c|c|c|c|}
\hline
Gate\rule[0.5ex]{0cm}{2ex} & $\lambda$ & $\eta_{4}$ & $Tr\, P$ &
  $\calF$\\
\hline
Hadamard\rule[0.5ex]{0cm}{2ex} & $5.8511$ & $2.9280\times 10^{-4}$ & 
  $8.82\times 10^{-6}$ & $0.9999\, 98$\\
NOT & $7.3205$ & $2.9277\times 10^{-4}$ & $1.10\times 10^{-5}$ &
  $0.9999\, 97$\\
Modified $\pi /8$ & $6.0150$ & $8.1464\times 10^{-4}$ & $3.03\times 10^{-5}$ 
   & $0.9999\, 92$\\
Modified phase & $5.9750$ & $3.8060\times 10^{-4}$ & $8.20\times 10^{-5}$ 
  & $0.9999\, 80$\\
\hline
\end{tabular}
}
\end{table}
presents the values for the sweep parameters $\lambda$ and $\eta_{4}$ that
produced our best results for $Tr\, P$ for each of the one-qubit gates in 
$\calGU$. In all one-qubit simulations, the dimensionless inversion time was
$\tau_{0}=80.000$. Since $P_{e}\leq Tr\, P$, we see that $P_{e}<10^{-4}$ 
for all one-qubit gates in $\calGU$. Table~\ref{table1} also gives the fidelity 
$\calF$ for each gate which is obtained from $Tr\, P$ via Eq.~(\ref{fidtrP}). 
We will discuss the robustness of these gates in Section~\ref{sec5}.

\subsection{Modified Controlled-phase Gate} 
\label{sec4.2} 
\noindent We complete the universal set $\calGU$ by presenting our simulation
results for the $\calG$-symmetrized TRP implementation of the modified 
controlled-phase gate $V_{cp}$. In the two-qubit computational basis
(eigenstates of $\sigma_{z}^{1}\sigma_{z}^{2}$),
\begin{equation}
V_{cp} = \left(  \begin{array}{rrrr}
                             1 & 0 &\phantom00 & 0\\
0 & 1 &\phantom00 & 0\\
0 & 0 & -1 & 0\\
0 & 0 &\phantom00 & 1
                         \end{array}
               \right) .
\label{Vcp2}
\end{equation}
TRP implementation of $V_{cp}$ \textit{without\/} symmetrized
evolution was reported in Ref.~\cite{lhg2}. The results presented 
there are superceded by the $\calG$-symmetrized TRP results presented
here. 

For purposes of later discussion, note that the parameters appearing 
in $\calH_{2}(\tau )$ (see Eq.~(\ref{twoqbtHam})) fall into two sets. 
The first set consists of the TRP sweep parameters $(\lambda,\eta_{4},
\tau_{0})$, while the second set $(c_{4},d_{1},\ldots , d_{4})$ consists
of system parameters for degeneracy-breaking, detuning, and coupling. We
partitioned the TRP sweep into $\calN=2500$ pulses sequences, with each 
sequence based on the four element symmetry group $\calG$ for $V_{cp}$ 
introduced in Section~\ref{sec3.3}. The optimized parameter values 
$\lambda = 5.04$, $\eta_{4}=3.0\times 10^{-4}$, $\tau_{0}=120.00$, 
$c_{4}=2.173$, $d_{1}=99.3$, $d_{2}=0.0$, $d_{3}=-0.41$, and 
$d_{4}=0.8347$ produced the following two-qubit gate $U_{a}$:
\begin{equation}
Re\,\left( U_{a}\right) = 
   \left(  \begin{array}{rrrr}
                  0.9999\, 98 & -0.0000\, 03 & -0.0000\, 15 & -0.0000\, 14\\
 0.0000\, 03 & 0.9999\, 97 & 0.0000\, 36 & 0.0002\, 61\\
 -0.0000\, 15 & 0.0000\, 34 & -0.9999\, 80 & -0.0038\, 18\\
 -0.0000\, 14 & -0.0002\, 57 & -0.0038\, 38 & 0.9999\, 81
                                                  \end{array}
                                        \right) ;
\label{ReVcp}
\end{equation}
\begin{equation}
Im\,\left( U_{a}\right) = 
   \left(  \begin{array}{rrrr}
                  -0.0021\, 51 & 0.0000\, 03 & -0.0000\, 10 & -0.0000\, 73\\
 -0.0000\, 03 &- 0.0021\, 80 & 0.0001\, 40 & -0.0003\, 25\\
0.0000\, 10 & -0.0011\, 40 & 0.0017\, 02 & 0.0045\, 34\\
 -0.0000\, 73 & -0.0003\, 28 & -0.0045\, 21 & -0.0017\, 78
                                                  \end{array}
                                        \right) .
\label{ImVcp}
\end{equation}
From $U_{a}$ and $U_{t}=V_{cp}$, we find: (i)~$Tr\, P = 8.87\times10^{-5}$;
(ii)~gate fidelity $\calF_{cp} = 0.9999\, 89$; and (iii)~$P_{e}\leq 8.87\times
10^{-5}$. We see that by adding symmetrized evolution to a TRP sweep we
obtain an approximation to $V_{cp}$ with $P_{e}< 10^{-4}$.

We see that it has been possible to use TRP sweeps to produce a high-fidelity
universal set of quantum gates $\calGU$, with each gate error probability
falling below the rough-and-ready estimate for the accuracy threshold for 
fault-tolerant quantum computing: $P_{e}<10^{-4}$. 

\section{Discussion} 
\label{sec5} 

\noindent We have presented simulation results which suggest that TRP sweeps 
should be capable of implementing the universal quantum gate set $\calGU$ 
 non-adiabatically and with gate error probabilities satisfying $P_{e}<
10^{-4}$. It is worth noting that all gates in $\calGU$ are driven by
a \textit{single\/} type of control field (TRP), and that the gate error 
probability for all gates in $\calGU$ falls below the rough-and-ready estimate
of the accuracy threshold $P_{a}\sim 10^{-4}$. These results suggest that 
TRP sweeps show promise for use in a fault-tolerant scheme of quantum 
computing. 

To achieve this high level of performance in our current
formulation of TRP, some of the TRP parameters must be controlled to high 
precision. For the one-qubit gates \cite{lhg1}, the critical parameter is 
$\eta_{4}$ which must be controlled to five significant figures to achieve
best gate performance. For the modified controlled-phase gate $V_{cp}$, the
critical parameters are \textit{not\/} the TRP sweep parameters. Instead, for
$V_{cp}$ \textit{without\/} symmetrized evolution \cite{lhg2}, the critical 
parameters are $c_{4}$, $d_{1}$, and $d_{4}$ which also require five 
significant figure precision. However, \textit{when group-symmetrized evolution
is added\/}, not only is TRP able to make an approximate $V_{cp}$ with $P_{e}<
10^{-4}$, but gate robustness is also \textit{improved\/}. Specifically, 
$d_{1}$ ceases to be a critical parameter, and $c_{4}$ and $d_{4}$ now 
only need to be controllable to four significant figures. Table~\ref{table2} 
shows how $Tr\, P$
\begin{table}[htb]
\tcaption{\label{table2}Sensitivity of $Tr\, P$ to small variation of $c_{4}$
and $d_{4}$ for the two-qubit gate $V_{cp}$. All other parameter values are 
as given in the text.}
\centerline{\footnotesize\smalllineskip
\begin{tabular}{|cc|cc|}
\hline
$c_{4}$\rule[0.5ex]{0cm}{2ex} & $Tr\, P$ & $d_{4}$ & $Tr\, P$ \\
\hline
$2.172$\rule[0.5ex]{0cm}{2ex} & $6.79\times 10^{-3}$ & $0.8346$ & 
  $1.52\times 10^{-3}$ \\
$2.173$ & $8.87\times 10^{-5}$ & $0.8347$ & $8.87\times 10^{-5}$ \\
$2.174$ & $7.73\times 10^{-3}$ & $0.8348$ & $1.52\times 10^{-3}$ \\
\hline
\end{tabular}
}
\end{table}
varies when either $c_{4}$ or $d_{4}$ is varied in the fourth significant figure, 
with all other parameters held fixed. Thus, adding group-symmetrized evolution 
improves both the accuracy and robustness of the TRP approximation to 
$V_{cp}$. Note that four significant figure precision corresponds to 14-bit
precision which can be realized with present-day arbitrary waveform generators
(AWG) \cite{tek}. On the other hand, the current precision requirements for all 
one-qubit gates in $\calG_{U}$ lie beyond the reach of existing commercially 
available AWGs. Unfortunately, group-symmetrized evolution cannot be used 
to improve the robustness of the one-qubit TRP gates. It is possible to show 
that if $U_{t}=\mathbf{a}\cdot\bfsig$, the only one-qubit unitary operators 
that commute with $U_{t}$ are the identity and a multiple of $U_{t}$. Thus 
the only symmetry group available that does not include $U_{t}$ is the trivial
group whose sole member is the identity. Some other means must be
found to improve the robustness of the TRP approximations to the one-qubit
gates in $\calGU$.  Two approaches are currently under study based on:
(i)~the Hessian of our cost function $Tr\, P$; and (ii)~quantum optimal control
theory. 

In previous work \cite{lgm} we have studied a number of forms of polynomial,
as well as periodic, twist. To date we have found that quartic twist provides
best all-around performance when it comes to making the gates in $\calG_{U}$.
Although we do not at present have arguments that explain why this is so,
ongoing work based on quantum optimal control provides a framework with which
this question can be studied. This represents an important direction for future
work.

Finally, Refs.~\cite{lhg2}--\cite{lhg1} have shown how TRP sweeps can be 
applied to NMR, atomic, and superconducting qubits. We note that 
TRP-generated quantum gates should also be applicable to spin-based qubits 
in quantum dots as such qubits also Zeeman-couple to a magnetic field.

\nonumsection{Acknowledgements} 
\noindent We thank: (i)~F. Nori, RIKEN, and CREST for making our visit to 
RIKEN possible;(ii)~RIKEN for access to the RIKEN Super Combined Cluster on 
which the simulations incorporating group-symmetrized evolution were done; 
and (iii)~T. Howell III for continued support.

\nonumsection{References} 


\noindent

\begin{thebibliography}{99} 
\bibitem{ft1} D. Aharanov and M. Ben-Or, \textit{Fault-tolerant computation 
with constant error\/}, in Proceedings of the Twenty-Ninth ACM Symposium on 
the Theory of Computing, 176 (1997).
\bibitem{ft2} A. Y. Kitaev, \textit{Quantum computation algorithms and 
error correction\/}, Russ.\ Math.\ Surv.\ \textbf{52}, 1191 (1997). 
\bibitem{ft3} A. Y. Kitaev, \textit{Quantum error correction with imperfect 
gates\/}, in Quantum Communication, Computing, and Measurement (Plenum 
Press, New York, 1997), pp.~181--188. 
\bibitem{ft4} D. Gottesman, \textit{Stabilizer codes and quantum
error correction\/}, Ph.~D. thesis, California Institute of Technology,
Pasadena, CA (1997). 
\bibitem{ft5} E. Knill, R. Laflamme, and W. H. Zurek, 
\textit{Resilient quantum computation\/}, Science \textbf{279}, 342 (1998).
\bibitem{ft6} E. Knill, R. Laflamme, and W. H. Zurek, \textit{Resilient 
quantum computation: error models and thresholds\/}, Proc.\ R. Soc.\ Lond.\ 
A \textbf{454}, 365 (1998). 
\bibitem{ft7} J. Preskill, \textit{Reliable quantum computers\/}, Proc.\ R. 
Soc.\ Lond.\ A \textbf{454}, 385 (1998). 
\bibitem{ft8} F. Gaitan, \textit{Quantum error correction and fault-tolerant 
quantum computing\/} (CRC Press, Boca Raton, FL 2008). 
\bibitem{zw1} J. W. Zwanziger, S. P. Rucker, and G. C. Chingas, 
\textit{Measuring the geometric component of the transition probability in a 
two-level system\/}, Phys.\ Rev.\ A \textbf{43}, 3232 (1991). 
\bibitem{lhg2} R.~Li, M.~Hoover, and F.~Gaitan, \textit{High-fidelity universal
set of quantum gates using non-adiabatic rapid passage\/}, Quantum Info.\ 
Comp.\ \textbf{9} 290 (2009). 
\bibitem{fg1} F. Gaitan, \textit{Temporal interferometry: a mechanism for 
controlling qubit transitions during twisted rapid passage with a possible 
application to quantum computing\/}, Phys.\ Rev.\ A \textbf{68} 052314 
(2003). 
\bibitem{lhg1} R. Li, M. Hoover, and F. Gaitan, \textit{High-fidelity
single-qubit
gates using non-adiabatic rapid passage\/}, Quantum Info.\ Comp.\ 
\textbf{7} 594 (2007).
\bibitem{zwan2} J.~W.~Zwanziger, U. Werner-Zwanziger, and F. Gaitan,
\textit{Non-adiabatic rapid passage\/}, Chem.\ Phys.\ Lett.\ \textbf{375} 
429 (2003). \bibitem{zan} P. Zanardi, \textit{Symmetrizing evolutions\/}, 
Phys.\ Lett.\ A \textbf{258} 77 (1999). 
\bibitem{abra} A. Abragam, \textit{Principles of nuclear magnetism\/} 
(Oxford University Press, New York 1961). 
\bibitem{fg2}F. Gaitan, \textit{Berry's phase in the presence of a
non-adiabatic environment with an application to magnetic resonance\/}, 
J. Mag.\ Reson.\ \textbf{139} 152(1999). 
\bibitem{f&h} R. P. Feynman and A. R. Hibbs, \textit{Quantum mechanics
and path integrals\/}, (McGraw-Hill, New York, 1965). 
\bibitem{numrec} W.H. Press et al., \textit{Numerical Recipes\/} (Cambridge
University Press, New York 1992).
\bibitem{vio&llyd} L. Viola and S. Lloyd, \textit{Dynamic suppression of 
decoherence in two-state quantum systems\/}, Phys.\ Rev.\ A \textbf{58}, 
2733 (1998).
\bibitem{tek} For example, the Tektronix AWG5000B arbitrary waveform
generator provides 14-bit vertical resolution.
\bibitem{lgm} R. Li and F. Gaitan, \textit{Controlling qubit transitions 
through quantum interference during non-adiabatic rapid passage},
Optics and Spectroscopy \textbf{99}, 257 (2005).
\end{thebibliography}
\end{document}